\documentclass[12pt]{article}
\usepackage{amssymb}
\usepackage{latexsym}
\usepackage{amsmath}
\usepackage{color}
\usepackage{cite}
\usepackage{units}
\usepackage{tabularx}
\usepackage[pdftex]{graphicx}

\numberwithin{equation}{section}

\newcommand{\lsim}{\raisebox{-0.13cm}{~\shortstack{$<$ \\[-0.07cm] $\sim$}}~}

\begin{document}
\pagestyle{empty}
\begin{flushright}
September 2010
\end{flushright}

\begin{center}
{\large\sc {\bf Hunting for CDF Multi--Muon ``Ghost'' Events at Collider and
    Fixed--Target Experiments}}  

\vspace{1cm}
{\sc Nicki Bornhauser$^{1}$ and Manuel Drees$^{1}$}  

\vspace*{5mm}
{}$^1${\it Physikalisches Institut and Bethe Center for Theoretical
  Physics, Universit\"at Bonn, Nussallee 12, D53115 Bonn, Germany} \\ 
\end{center}

\vspace*{1cm}
\begin{abstract}
  In 2008 the CDF collaboration discovered a large excess of events
  containing two or more muons, at least one of which seemed to have
  been produced outside the beam pipe.  We investigate whether
  similar ``ghost'' events could (and should) have been seen in
  already completed experiments. The CDF di--muon data can be
  reproduced by a simple model where a relatively light $X$ particle
  undergoes four--body decay. This model predicts a large number of
  ghost events in Fermilab fixed--target experiments E772, E789 and
  E866, applying the cuts optimized for analyses of Drell--Yan
  events. A correct description of events with more than two muons
  requires a more complicated model, where two $X$ particles are
  produced from a very broad resonance $Y$. This model can be tested
  in fixed--target experiments only if the cut on the angles, or
  rapidities, of the muons can be relaxed. Either way, the UA1
  experiment at the CERN $p \bar p$ collider should have observed
  ${\cal O}(100)$ ghost events.
\end{abstract}

\newpage
\setcounter{page}{1}
\pagestyle{plain}

\section{Introduction}

In 2008 the CDF Collaboration published an analysis of events
containing at least two muons, and found a large excess of
so--called ghost events, which supposedly cannot be explained
by the known QCD production with the current understanding of the CDF
detector \cite{Ghost}. The two muons with the highest transverse
momenta in a sample event, the so-called initial muons, each were
required to have transverse momentum $p_T \geq \unit[3]{GeV}$, a
pseudorapidity $|\eta| \leq 0.7$ and a combined invariant mass in
the range $\unit[5]{GeV} < m_{\mu\mu} \leq \unit[80]{GeV}$. In
the following these cuts are called two--muon cuts. 

In defining these muon tracks, no information from the silicon
microvertex detector (SVX) is used. According to CDF, some
\unit[24]{\%} of muon pairs passing the above cuts should have been
detected by the SVX if they come from known sources. These are called
QCD background events by CDF, and chiefly originate from Drell--Yan
pairs, semileptonic decays of $c$ and $b$ quarks, and misidentified
charged hadrons. However, only in about \unit[19]{\%} of the observed
di--muon events both muons were also detected by the SVX. CDF
therefore concludes that there is a large number of ghost events where
at least one initial muon is produced outside of the beam pipe (more
exactly, outside the first layer of the SVX, which is adjacent to the
beam pipe) with a radius of $\unit[15]{mm}$. Indeed, many of the
primary muons were found to have a large impact parameter.

Moreover, nearly \unit[10]{\%} of these events contained one or several
additional muons with $p_T \geq \unit[2]{GeV}$ and $|\eta| \leq
1.1$, many of which again have a high impact parameter; this fraction
is about four times higher than one expects for QCD events.
Furthermore, the ghost sample contains approximately equally many
same--sign (SS) and opposite--sign (OS) initial muon pairs.

After correcting for events from ordinary sources, i.e. QCD
production, CDF finds $84895 \pm 4829$ ghost events within a data
sample corresponding to an integrated luminosity of
$\unit[742]{pb^{-1}}$. The ghost cross section
\begin{equation*}
\sigma(p\bar{p} \rightarrow {\rm ghosts}) = \unit[\frac{84895 \pm
  4829}{742}]{pb} \approx \unit[(114.4 \pm 6.5)]{pb} 
\end{equation*}
is comparable in size to the $b\bar{b}$ contribution to the di--muon sample:
\begin{equation*}
\sigma(p\bar{p} \rightarrow b\bar{b} \rightarrow \mu\mu) =
\unit[\frac{221564 \pm 11615}{742}]{pb} \approx \unit[(298.6 \pm
15.7)]{pb}. 
\end{equation*}

Considering this high cross section and the remarkable properties of
the ghost events it is natural to ask whether such events could (and
hence presumably should) have been observed in earlier
experiments. This question can be answered only in the framework of
concrete particle physics models. In the following Section we
therefore first describe properties of the ghost particles (whose
decays produce the detected muons) that can be derived almost
model--independently from the CDF data. In Sec.~\ref{sec:SM_Results}
we describe a simple model that reproduces most of the features of
most ghost events \cite{Ghost}. This model indeed predicts that
experiments at lower energies should have observed dozens to thousands
of ghost events. We then construct a somewhat more complicated model,
which improves the description of the subset of ghost events
containing at least three muons; this model is much more difficult to
probe at fixed--target experiments. Finally, we conclude.

\section{General Considerations}

As described above, muons in ghost events seem to originate at a large
distance from the primary interaction vertex. This indicates that the
muons are produced in the decays of rather long--lived $X$ particles.

Many properties of the $X$ particle follow from the properties of the
ghost events. First, the $X$ particles should be electrically
neutral. Otherwise they would have escaped detection by LEP
experiments only if their mass exceeded \unit[100]{GeV}, which would have put
their cross section closer to that for $t \bar t$ production than that
for $b \bar b$ production. Moreover, the $X$ particles themselves
would then have produced a track which in many cases should have been
easier to detect by the SVX than a muon track, since such heavy
$X$ particles would often have had smaller velocity, and hence larger
energy loss $dE/dx$, than ultra--relativistic muons.

The $X$ particles should have an average decay length $\gamma \tau_X v
\geq \unit[15]{mm}$, in order to account for the high impact
parameters of the ghost muons and in particular for the fact that at
least one initial muon appears to be created outside of the beam
pipe. On the other hand, the decay length cannot be very much larger
than 15 mm, since both muons originate well within the CDF tracker.

The fact that approximately equally many SS and OS ghost di--muon
events are observed indicates that the $X$ particle should be a
Majorana particle, i.e. identical to its CP conjugate.

Finally, the fact that a significant fraction of ghost di--muon events
contains additional ``secondary'' muon indicates that all $X$
particles decay into a final state with relatively high multiplicity.
Otherwise the branching ratio into multiple muons would be expected to
be very small. On the other hand, the higher the decay multiplicity,
the more complicated a full (renormalizable) quantum field theory
reproducing our phenomenological model would have to be. In our
simulations we therefore assume that all $X$ particles decay into four
elementary fermions, at least one of which is a muon. We model these
decays using pure phase space, i.e. assuming constant decay matrix
elements.

\section{Simple Model} \label{sec:SM_Results}

In order to proceed further, we have to make assumptions regarding the
production mechanism of $X$ pairs. In our simple model we assume that
the $X$ particles are pair--produced directly in either gluon--gluon
fusion or quark--antiquark annihilation, with differential $S-$wave
cross section
\begin{equation} \label{e1}
\frac{d\sigma(gg/q\bar{q} \rightarrow XX)} {d\cos \theta} =
N_{gg/q\bar{q}} \cdot \frac{\beta}{\hat{s}} = N_{gg/q\bar{q}} \cdot
\frac{ \sqrt{1 - \frac {4 m^2_X} {\hat s} } } {\hat s}. 
\end{equation}
Here $\hat{s}$ is the squared partonic center of mass energy, $m_X$ is
the mass of the $X$ particle, $\beta$ is the velocity of the $X$
particles in the partonic center of mass frame, and $N_{gg/q\bar{q}}$
are constants which are fixed by the requirement that we reproduce the
ghost cross section measured by CDF. 

We want to use the simple model to estimate the number of ghost
di--muon events that should have been detected by earlier experiments
operating at lower center of mass energies. Using the fact that about
half of these events contain like--sign muons, as well as the large
impact parameters of these muons, should suppress physics backgrounds
to negligible levels. Possible instrumental backgrounds can only be
evaluated by the experiments themselves. 

The free parameters of this simple model are the decay modes and
corresponding branching ratios of the $X$ particle, its mass and
lifetime. We set the lifetime as
\begin{equation}\label{equ:cTau_X}
c \tau_X = \unit[20]{mm}\,;
\end{equation}
this choice only affects the impact parameter distribution of the
ghost muons, but no other results.

We include the following decay modes:
\begin{itemize}
\item 1--muon: $X \rightarrow \mu^- \bar{\nu}_{\mu} u \bar{d}$ \ or \
  $X \rightarrow \mu^+ \nu_{\mu} \bar{u} d$ 
\item 2--muon: $X \rightarrow \mu^- \mu^+ u \bar{u} $ \ or \ $X
  \rightarrow \mu^- \mu^+ d \bar{d}$ 
\item 4--muon: $X \rightarrow \mu^- \mu^+ \mu^- \mu^+$
\end{itemize}
Our assumption the $X$ is a Majorana particle implies that charge
conjugate modes contribute with equal branching ratio. These decays
conserve electric charge and all lepton numbers. Spin conservation
then implies that $X$ is a boson.

In order to simulate $p \bar p \rightarrow X X \rightarrow \mu^+ \mu^-
+ \dots$ events, we implemented the production cross sections
(\ref{e1}) into HERWIG++ \cite{Herwig}, using default parameters for
parton showering and the underlying event. Recall that we model $X$
decays assuming constant decay matrix elements.

Fig.~5 in Ref.~\cite{Ghost} shows that the di--muon excess occurs at
rather small invariant mass of the primary muon pair, $m_{\mu\mu}
\lsim \unit[40]{GeV}$. Together with the large cross section for ghost
events this argues for a relatively light $X$ particle. Unfortunately
no $p_T$ spectrum of the primary muons in ghost events is provided,
which might have allowed to estimate $m_X$ from the bulk of events
that do not contain additional muons.

\begin{figure}[htb]
  \centering
  \includegraphics[width=17cm]{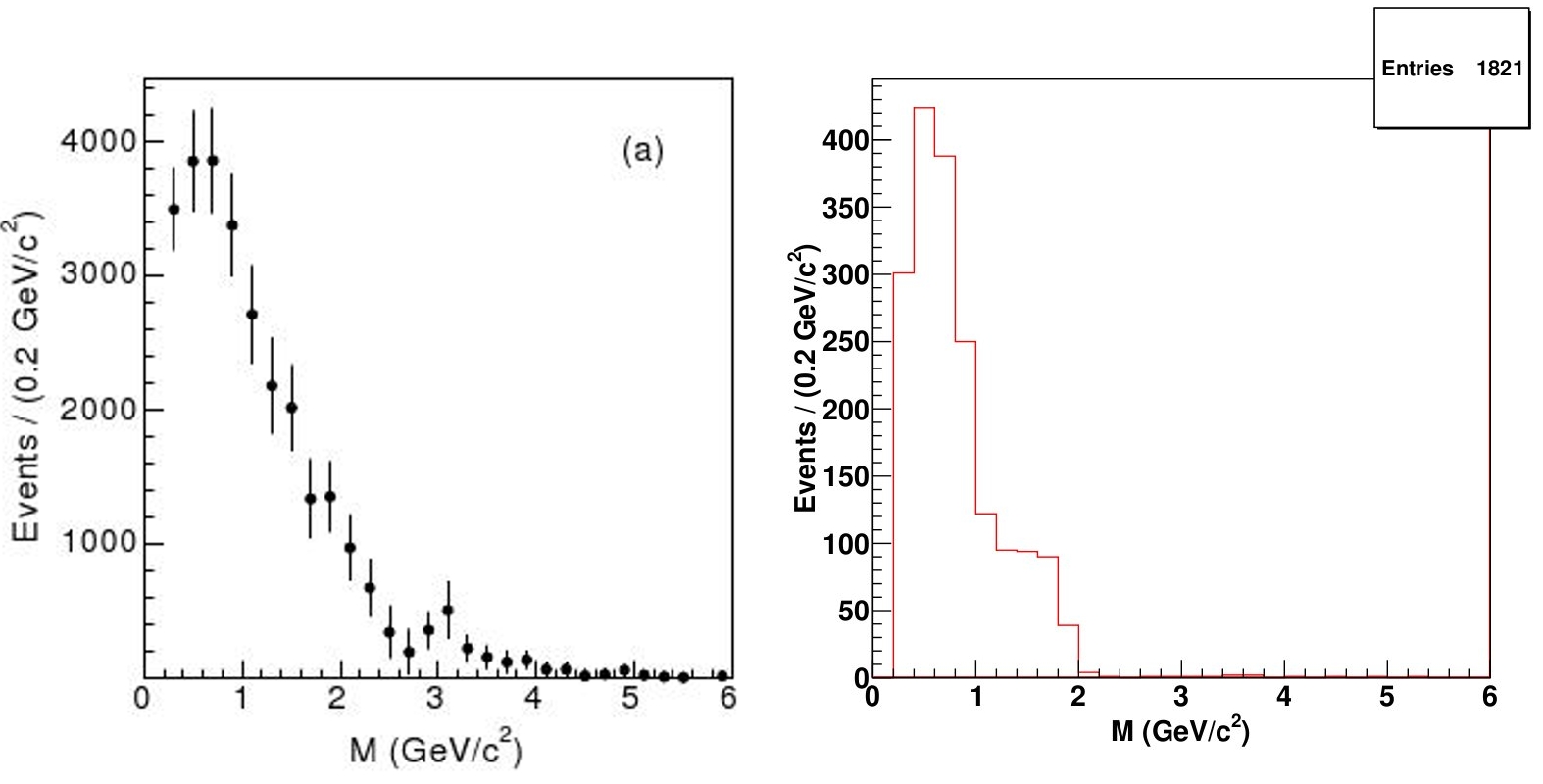}	
\caption[]{Invariant mass distribution of all muons within
  $36.8^\circ$ cones around initial muons which contain at least one
  additional muon, for the ghost measurement (\textbf{left}) [Fig.~34a
  in \cite{Ghost}] and our simulation (\textbf{right}). The simple
  model parameters for the process $q \bar q \rightarrow XX$ are
  used in the simulation. 21,745 of 40 million simulated events pass
  the two--muon cuts.} 
  \label{fig:Fig34a}
\end{figure}

We therefore use the invariant mass distribution of all muons
contained in the 27,990 $36.8^\circ$ cones (corresponding to $\cos
\theta = 0.8$) around initial muons which contain at least one
additional muon (see Fig.~34a in \cite{Ghost}).\footnote{Note that the
  analysis of events with more than two muons uses a larger data
  sample ($\int {\cal L} dt = \unit[2100]{pb^{-1}}$) than that used
  for the determination of the ghost cross section described in the
  Introduction. The number of cones given here is consistent with our
  earlier statement that a little under \unit[10]{\%} of all ghost
  events contain at least one additional muon.} This distribution
matches the invariant mass distribution in each cone for the events,
in which both cones contain at least one additional muon (Fig.~34b in
\cite{Ghost}). This agrees with our model, in which both $X$ particles
contribute one initial muon each with identical cone properties. In
most cases only additional muons from the decay of the same $X$
particle that produced the initial muon are contained in the
$36.8^\circ$ cone around the initial muon. The reason is that in the
partonic center of mass system the $X$ particles propagate in opposite
directions. The measured invariant mass distribution of all muons in
these cones is shown in the left frame of Fig.~\ref{fig:Fig34a}; it
peaks around $\unit[0.6]{GeV}$, and becomes very small beyond
$\unit[3]{GeV}$. Our best fit value for the mass of the $X$ particle
is
\begin{equation}\label{equ:m_X}
m_X = \unit[1.8]{GeV}.
\end{equation}
The corresponding distribution is shown in the right frame of
Fig.~\ref{fig:Fig34a}, ignoring measurement errors and assuming $X$
pair production from $q \bar q$ annihilation. We consider the
agreement satisfactory.

Finally, we determine the branching ratios for the various $X$ decay
modes by using the sign--coded muon multiplicity distribution of
additional muons found in $36.8^\circ$ cones around the initial
muons. Here each additional muon with the {\em same charge} as the
primary muon increases the count by ten, whereas each additional muon
with the {\em opposite charge} as the primary muon counts as one. In
case of $X$ pair production from quark--antiquark annihilation our fit
for the branching ratios is \unit[93.88]{\%} for the 1--muon,
\unit[5.02]{\%} for the 2--muon, and \unit[1.10]{\%} for the 4--muon
decay. The gluon fusion mechanism requires slightly different
branching ratios, because of the different efficiencies for the
generated muons to pass the cuts.

\begin{figure}[htb]
  \centering
  \includegraphics[width=17cm]{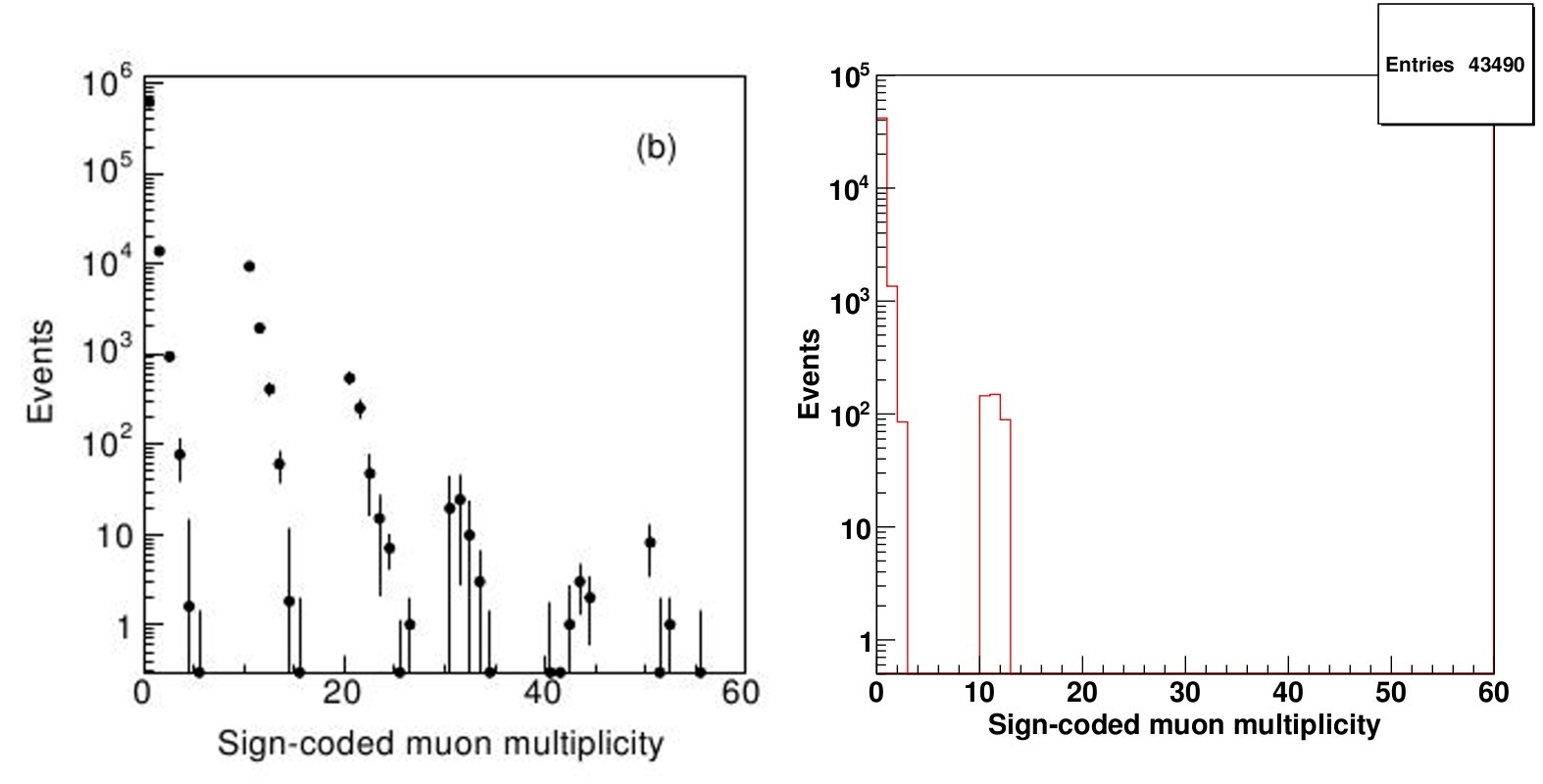}
    \caption[]{Sign-coded muon multiplicity distribution of additional
      muons found in $36.8^\circ$ cones around initial muons for the ghost
      measurement (\textbf{left}) [Fig.~22b in \cite{Ghost}] and our
      simulation (\textbf{right}). The simple model parameters for the
      process $q\bar{q} \rightarrow XX$ are used in the
      simulation. 21,745 of 40 million simulated events pass the
      two--muon cuts.}
  \label{fig:Fig22b}
\end{figure}

Fig.~\ref{fig:Fig22b} compares the original CDF result (Fig.~22b in
\cite{Ghost}) with our simulation. Clearly our model produces fewer
entries at high muon multiplicities. Furthermore, the ratio between
cones with one OS additional muon, i.e. the number of entries in
``1'', and with one SS additional muon, i.e. the number of entries in
``10'', is too high. The higher number of entries in ``1'' follows
from the construction of the decay modes. Since all muons inside a
cone usually result from the decay of the same $X$ particle, the 2--muon
decay only contributes to ``1'' and the 4--muon decay gives two times
more entries in ``1'' than in ``10''. In order to reproduce this ratio
correctly, we need a 2--muon decay with SS muons. The conservation of
electric charge, all single lepton numbers and spin then requires a
decay into at least eight elementary fermions, e.g. $X \rightarrow
\mu^+ \mu^+ \nu_{\mu} \nu_{\mu} \bar{u} d \bar{u} d$. One would
obviously need a very contrived model to reproduce this in a
renormalizable quantum field theory. Another possibility is to allow
violation of separate lepton numbers, insisting only that the {\em
  total} lepton number remains conserved; we will pursue this in our
more complicated model.

Furthermore, our simple model almost never generates events with more
than three additional muons in a $36.8^\circ$ cone around a primary
muon. This discrepancy with the data may be considered less
problematic, since it concerns only a small fraction of all ghost
events. In addition these multi--muon events are presumably more
prone to experimental errors. For example, an underestimated fake muon
rate would affect the extracted rate of events with multiple muons
more than that of events with fewer muons; note that the fake muon
contribution has been subtracted by CDF in the result reproduced in
the left frame of Fig.~\ref{fig:Fig22b}.

We emphasize that we want to use our model only to estimate di--muon
rates. While the multi--muon events are indeed spectacular, they have
substantially reduced cross sections. Recall also that there is
very little physics background to di--muon events where the muons
originate a few cm from the primary interaction vertex.\footnote{Known
  hadron decays yielding muons occur either much earlier (for $c$ or
  $b$ flavored hadrons), or typically have much longer decay lengths
  (e.g. $c \tau = \unit[7.8]{m}~[\unit[3.7]{m}]$ for charged pions [kaons], which
  needs to be multiplied with a large $\gamma$ factor to produce
  sufficiently hard muons).}

Having fixed the parameters of the model, we are ready to make
predictions for the number of expected ghost events in various
experiments. We focus on experiments that had the possibility to
identify muons, and accumulated large data samples. These include the
experiment UA1 at the CERN $p \bar p$ collider \cite{UA1}, and the
Fermilab fixed--target experiments E605 \cite{E605}, E772 \cite{E772},
E789 \cite{E789} and E866 \cite{E866}. E789 is especially interesting
for our purposes since it featured a vertex detector, which should
have had no trouble detecting the long flight path of $X$ particles if
some of them had been produced. We also considered the HERA
experiments ZEUS and H1, where $X$ pairs could have been produced in
``resolved'' photoproduction processes involving the parton
distribution functions inside the photon; however, we found that they
are not sensitive to our model, since the cross section is suppressed
by a factor $\alpha_{\rm em}^2$ and the integrated luminosity is not
very high.

Table~\ref{tab:DataSets} shows the cuts which we use for the
simulations. Since we focus on di--muon events, we only cut on the
momenta of these muons. In case of the UA1 experiment, the cuts can be
taken directly from their analyses of di--muon data \cite{UA1}. The
fixed--target experiments are more difficult to simulate, since not
enough information is provided in their analyses of di--muon data to
completely determine their acceptance region. It should be noted that
these analyses were optimized for Drell--Yan muon pairs, which to
leading order emerge back--to--back in the partonic center of mass
system. This is usually not the case for the muon pairs from $X$
decays.

\begin{table}
\centering
\caption[Data sets]{Cuts on the two muons in the 
  experiments we consider. Except for E789 all cuts are applied in the
  hadronic center of mass system. The cuts for E789 are applied in the
  lab frame.}
\begin{tabular}{|l|l|l|l|l|}
\hline
Exp. 	& $ \sqrt{s} $ [GeV]& $p_{T,\,{\rm min}}$ [GeV] & Inv. mass [GeV]
& Pseudorapidity \\ \hline 
CDF 	& 1960 	& 3 & $ 5 < m_{\mu\mu} \leq 80 $ & $ |\eta| \leq 0.7 $\\  
UA1--a 	& 546 	& 3 & $ 6 < m_{\mu\mu} $ & $ |\eta| \leq 0.7 $\\
UA1--b 	& 630 	& 3 & $ 6 < m_{\mu\mu} $ & $ |\eta| \leq 0.7 $\\
UA1--c 	& 630 	& 3 & $ 6 < m_{\mu\mu} < 35 $ & $ |\eta|
\leq 0.7 $\\ 
E605--a 	& 38.8 	& --- & $ 7 < m_{\mu\mu} < 18 $ & $ |\eta| \leq 0.02 $\\ 
E605--b 	& 38.8 	& --- & $ 6 < m_{\mu\mu} < 18 $ & $ |\eta| \leq 0.02 $\\ 
E772--a 	& 38.8 	& --- & $ 4.5 \leq m_{\mu\mu} \leq 9$ or $11 \leq
m_{\mu\mu} $ & $ |\eta| \leq 0.044 $\\  
E772--b 	& 38.8 	& --- & $ 4 \leq m_{\mu\mu} $ & $ |\eta| \leq 0.044 $\\ 
E789 	& 38.8 	& --- & $ 2 \leq m_{\mu\mu} \leq 6$ & $3.506 \leq \eta
\leq 4.605$ \\ 
E866 	& 38.8 	& --- & $ 4 \leq m_{\mu\mu} \leq 9 $ or $ 10.7 \leq
m_{\mu\mu} $ & $ |\eta| \leq 0.02 $\\
\hline
\end{tabular}
\label{tab:DataSets}
\end{table}

In detail, the fixed--target experiments do not cut on $p_T$, since
they are sensitive to muons even in the very forward direction. The
lower bounds on the di--muon invariant mass $m_{\mu\mu}$ are
determined from the acceptances of these experiments as described in
their publications. The upper bounds on $m_{\mu\mu}$ are basically
irrelevant given the small center of mass energy of these
experiments. Finally, the (very stringent!) cut on the pseudorapidity
$|\eta|$ results from the main focus on Drell--Yan events within these
experiments. In the partonic center of mass system the angular
dependence of a created Drell--Yan di--muon pair is given by $(1 +
\cos \theta_{cs})$. For E605 the acceptance of the muon--detectors is
restricted to a ``small range of the decay angle'' near $ \theta_{cs}
= 90^\circ$ \cite{E605}. Because this small range is not further
specified, we use the acceptance plots in Figs.~8 and 9 in
Ref.~\cite{E605} to estimate a spatial coverage of around
\unit[2]{\%}, i.e. $\theta_{cs} \in [-0.02, 0.02]$. In the limit where
the angular distributions of both muons are independent of each other,
the acceptance decreases to $\unit[0.04]{\%}$, resulting in a much
worse acceptance for ghost events than for Drell--Yan events.
Moreover, in contrast to Drell--Yan events we cannot reconstruct the
partonic center of mass system from the measured primary muons in
ghost events. We therefore work within the hadronic center of mass
system. The pseudorapidity cut $|\eta| \leq 0.02$ given in
Table~\ref{tab:DataSets} then approximately reproduces the angular
coverage given above.

Like E605 the fixed--target experiment E772 features a small spatial
muon coverage which is not specified clearly \cite{E772}. Therefore we
again have to estimate the pseudorapidity cut. We compared the data
sets E605--a, E605--b and E772--a, taking into account their different
di--muon invariant mass acceptances and integrated luminosities. Using
the given differential Drell--Yan cross section $m^3_{\mu^+\mu^-} d^2
\sigma^{DY} / dm_{\mu^+\mu^-} dx_F$ for fixed Feynman$-x$ we can
determine the invariant mass dependence of the total cross section;
see Fig.~3 in the first and Table~1 in the second publication of
Ref. \cite{E772}. We find $\sigma^{DY} \propto m^{-5}_{\mu^+\mu^-}$,
where $m_{\mu^+\mu^-}$ is the lower mass limit. Starting from the
\unit[2]{\%} acceptance of E605 we estimate the cut $|\eta| \leq 0.044
$ for E772; see Table~\ref{tab:DataSets}.

We use the pseudorapidity cut $|\eta| \leq 0.02$ for the fixed--target
experiment E866, because no further information on the muon acceptance
is provided in Ref.~\cite{E866}. Note that the detectors of
all the Fermilab fixed--target experiments we consider are based on the original E605 detector. The experiment E789 provides
explicit pseudorapidity coverage information within the lab frame
\cite{E789}.

The CDF efficiency for identifying a muon pair is approximately
$\unit[26]{\%}$ \cite{Ghost}. We adopt this efficiency for the UA1
experiment as well. The fixed--target experiments have efficiencies of
the order of $\unit[90]{\%}$ after cuts.

Our results are shown in Table~\ref{tab:ResultsSM}. The errors of the
predictions result from the errors of the CDF measurement, of the
simulated cross sections, of the integrated luminosities, of the
efficiencies for the di--muon acquisition and of the efficiency for
the simulated events to pass the di--muon cuts. The last of these
errors dominates for most of the fixed--target experiments. Since the
acceptance is very small, it is difficult to accumulate sufficient
statistics. We tried to overcome this problem by relaxing the upper
limit $\eta_{\rm max}$ on the absolute value of the pseudorapidity in
the hadronic center of mass frame, taking different values for this
cut. We then fit the resulting efficiency to a quadratic function of
$\eta_{\rm max}$, which we finally extrapolate to the value of
$\eta_{\rm max}$ given in Table~1.

\begin{table}
\centering
\caption[]{Expected number of ghost events for the simple model
  compared to the number of actually observed di--muon events. The
  fixed--target data sets only include OS di--muons. For E789 only the
  number of $J/\psi$ candidates is stated; the di--muon invariant mass
  distribution can be found in Fig.~1 in Ref.~\cite{E789}. The 
  normalization constants are $ N_{gg} \approx (7.1 \pm 0.4) \cdot 10^{-5}$
  and $ N_{q\bar{q}} \approx (1.8 \pm 0.1) \cdot 10^{-3}$. The branching ratios
  for the gluon fusion mechanism are $\unit[93.32]{\%}$ for the 1--muon,
  $\unit[5.6]{\%}$ for the 2--muon and $\unit[1.08]{\%}$ for the
  4--muon decay.} 
\begin{tabular}{|l|l|l|l|l|}\hline
 & & \multicolumn{2}{c|}{\# ghost events for} & \# of observed \\
Exp. & $\int {\cal L} dt$ [Nucl./pb] & $gg \rightarrow XX$
& $q\bar{q} \rightarrow XX$ & $\mu\mu$ events \\\hline 
CDF & $ 114.41 \pm 6.51 $ & $ 84895 \pm 4829 $ & $ 84895 \pm 4829$ & $
84895 \pm 4829$\\  
UA1--a & 0.108 & $ 1.5 \pm 0.1$ & $5.6 \pm 0.3$ & \multicolumn{1}{c|}{}\\
UA1--b & 0.6 & $10.2 \pm 0.6$ & $33.9 \pm 1.9$ & \raisebox{1.5ex}[-1.5ex]{880}\\
UA1--c & 4.7 & $79.5 \pm 4.5$ & $261.5 \pm 14.9$ & 2444\\
E605--a & $(1.14 \pm 0.08) \cdot 10^6$ & $30.4 \pm 10.2$ & $680.4 \pm
325.6$ & 43663 (OS)\\
E605--b & $(2.7 \pm 0.2) \cdot 10^5$ & $2.0 \pm 3.8$ & $165.3 \pm
151.2$ & 19470 (OS)\\
E772--a & $(5.8 \pm 0.3) \cdot 10^4$ & $39.9 \pm 2.3$ & $885.5 \pm
68.9$ & 83080 (OS)\\
E772--b & $3.5 \cdot 10^5$ & $251.7 \pm 28.8$ & $9415.1 \pm 547.1$ & $
\approx 450000 $ (OS)\\ 
E789 & $17.52 \pm 1.89$ & $134.7 \pm 16.5$ & $1294.9 \pm 158.7$ & $
71206 \pm 287 $ ($J/\psi$)\\ 
E866 & $3.78 \cdot 10^5$ & $144.4 \pm 39.7$ & $2339.2 \pm 135.9$ & $
\approx 360000 $ (OS)\\ 
\hline
\end{tabular}
\label{tab:ResultsSM}
\end{table}

We see that UA1 should have recorded about 100 (300) ghost di--muon
pairs if $X$ pairs are produced predominantly from gluon fusion ($q
\bar q$ annihilation). Recall that the cross section is normalized to
the CDF data. Since the gluon distribution function inside the proton
is softer than that of valence quarks, the cross section from gluon
fusion decreases faster with decreasing $\sqrt{s}$ than that from $q
\bar q$ annihilation. Note that UA1 did record about 3,300 di--muon
events with the cuts listed in Table~\ref{tab:DataSets}; about one
quarter of these events contained a same--sign muon pair
\cite{UA1}. The UA1 data are compatible with SM predictions, with the
biggest single contribution coming from $b \bar b$ production. It is
not clear to us whether the UA1 detector would have been able to
detect the rather long flight paths of the $X$ particles.\footnote{A
  vertex detector was briefly installed in UA1 at the end of the 1985
  run \cite{mueller}, but apparently was not used in the 1988/89
  running period.}

In case of the fixed--target experiments, we list the effective
luminosity for nucleon--nucleon collisions. Here the predicted number
of ghost events differs by about a factor of twenty between the two
production mechanisms, with gluon fusion again leading to fewer
events. However, even in that case we expect more than 100 events each
in experiments E772, E789 and E866. This is far smaller than the total
di--muon samples of these experiments. Recall, however, that in half
of the ghost events both muons have the same charge. Moreover, the
vertex detector of E789 should have been able to detect the displaced
vertices from the $X$ decay. Although these experiments did not (yet)
perform dedicated searches for ghost events, it seems unlikely that
they would have escaped detection within the samples of Drell--Yan
events.

\section{More Complicated Models}

\subsection{Breit-Wigner Resonance}\label{sec:BW}

\begin{figure}[htb]
  \centering
  \includegraphics[width=17cm]{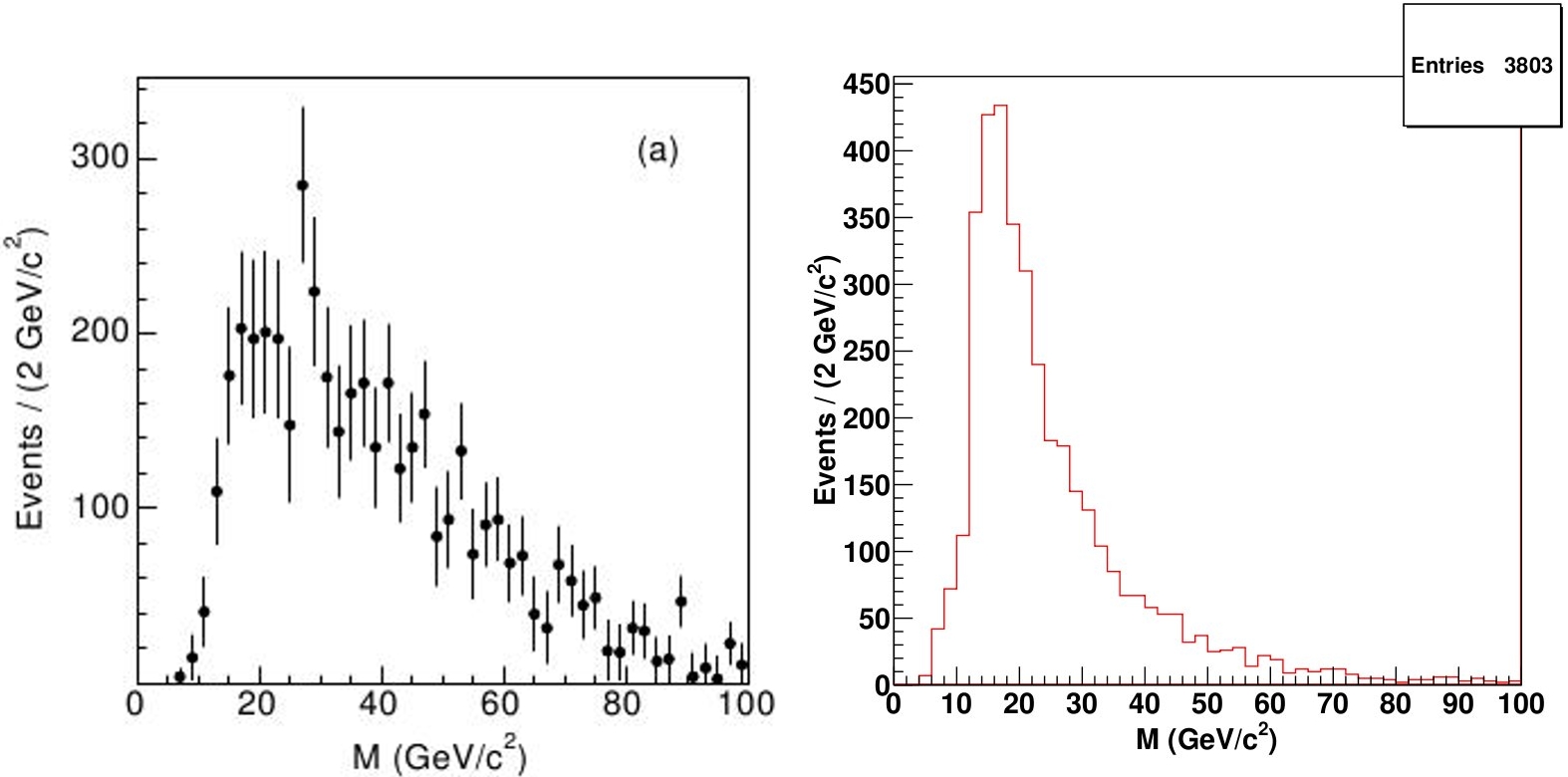}
    \caption[]{Invariant mass distribution of all muons within events
      in which both cones contain each at least one additional muon,
      for the ghost measurement (\textbf{left}) [Fig.~35a in
      \cite{Ghost}] and the simulation of our simple model
      (\textbf{right}). The process $ q\bar{q} \rightarrow XX $ is
      simulated. In order to improve the efficiency, we forced both
      $X$ particles to decay into two muons; the small four muon
      channel does not change the result significantly. 20,212 out of
      20 million generated events pass the cuts.}
  \label{fig:Fig35a}
\end{figure}

When fitting the parameters of the simple model discussed 
in the previous Section, we only compared to distributions of muons within
a $36.8^\circ$ cone around the primary muons; see
Figs.~\ref{fig:Fig34a} and \ref{fig:Fig22b}. In  Fig.~\ref{fig:Fig35a}
we instead show the total invariant mass distribution of all muons in
the small subsample of events in which the cones around {\em both}
primary muons contain each at least one additional muon (Fig.~35a in
\cite{Ghost}); these events thus contain a total of at least four
muons. We see that our simple model predicts this distribution to peak
slightly too early, and to fall off towards higher invariant masses
much faster than the CDF data do. Recall that the invariant mass
distribution of muons {\em within} each cone forced us to chose a
rather small mass for the $X$ particle. We can thus only increase the
number of events with large invariant mass of the multi--muon system
by modifying the production cross section. 

\begin{figure}[htb]
\centering
\includegraphics[width=8cm]{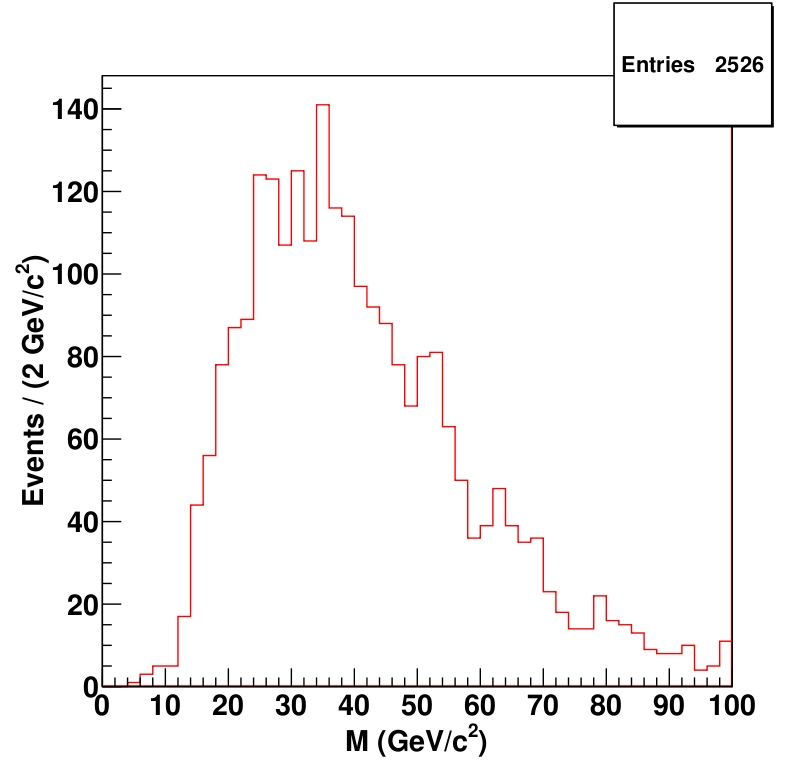}
\caption[]{Invariant mass distribution of all muons within events in
  which both cones contain each at least one additional muon,
  simulated assuming $q\bar{q} \rightarrow Y \rightarrow XX$. The
  branching ratios are $\unit[93.43]{\%}$ for  
  the 1--muon, $\unit[4.7]{\%}$ for the 2--muon and $\unit[1.87]{\%}$
  for the 4--muon decay. 807,162 of 20 million generated events pass
  the two--muon cuts.}
  \label{fig:Fig35aBWqqbar}
\end{figure}

To this end, we introduce a Breit--Wigner (resonance) factor in the
differential cross section:
\begin{equation}
\frac{d\sigma(gg/q\bar{q} \rightarrow Y \rightarrow XX)} {d\cos
  \theta} = N_{gg/q\bar{q}}^{BW} \cdot \frac{\hat{s}^2}{(\hat{s} -
  m_Y^2)^2 + \Gamma_Y^2 m_Y^2} \cdot \frac{\sqrt{1 - \frac{4
      m^2_X}{\hat{s}}}}{\hat{s}} 
\end{equation}
The constants $N_{gg/q\bar{q}}^{BW}$ are again chosen such that the
total cross section measured by CDF is reproduced.  $m_Y$ and
$\Gamma_Y$ are the mass and width of the resonance.
Fig.~\ref{fig:Fig35aBWqqbar} shows that we can reproduce the data
assuming $X$ pair production from quark--antiquark annihilation with
$m_Y = \Gamma_Y = \unit[50]{GeV}$. Note that we can reproduce the
slow fall--off towards high invariant masses only if the width has the
same order of magnitude as the mass of the resonance, which implies
that it is strongly coupled. The success of standard QCD in describing
jet data at a variety of colliders suggests that the coupling of $Y$
to quarks or gluons is not very large. On the other hand, the coupling
of $Y$ to $X$ is not constrained. In this scenario we expect most $Y$
``particles'' to ``decay'' into $X$ pairs. 

Since many $X$ particles are now produced with sizable transverse
momentum, the efficiency for passing the di--muon cuts is larger than
for the simple model. The muons produced in the two-- and four--muon
decay modes of $X$ are also somewhat more likely to pass the cut on
additional muons. The branching ratios of $X$ therefore have to be
chosen slightly differently from the simple model.

In this model we do not expect any ghost events in the fixed--target
experiments we analyzed after the cuts of Table~\ref{tab:DataSets}
have been imposed. This follows from two effects, which result from
the higher center of mass energy of CDF ($\unit[1.96]{TeV}$) compared
to the fixed--target experiments ($\unit[38.8]{GeV}$). 

First, for given normalization, introducing the Breit--Wigner factor
reduces the cross section.\footnote{The Breit--Wigner factor increases
  the partonic cross section for $\hat s > (m_Y^2 +
  \Gamma_Y^2)/2$. However, for $\Gamma_Y = m_Y$ it increases the
  partonic cross section by at most a factor of two. This cannot
  compensate the much stronger suppression of the partonic cross
  section at $\hat s \ll m_Y^2$, where the parton fluxes are much
  higher.} This reduction factor is much larger at the fixed--target
experiments operating at $\sqrt{s} < m_Y$. For fixed ghost cross
section at CDF, this reduces the fixed--target cross section before
cuts by a factor of 64 (35) for the quark--antiquark annihilation
(gluon fusion) mechanism.

Second, as mentioned above, introducing the Breit--Wigner factor
increases the cut efficiency for the CDF experiment. This effect is
much smaller for the fixed--target experiments, where the parton
densities force most events to still have small partonic center of
mass energy. Since we normalize to the CDF cross section after cuts,
this reduces the event rate at the fixed--target experiments, e.g. for
E789, by another factor of 16 for quark--antiquark annihilation and 41
for gluon fusion. In combination, these two effects reduce the
expected event number for the fixed--target experiments by three
orders of magnitude.

\subsection{Muon Number Violating $X$ Decays}\label{sec:tauDecay}

We saw in Sec. \ref{sec:SM_Results} that our simple model
does not reproduce the sign--coded multiplicity distribution of
additional muons very well. We argued that this is inevitable unless
we allow $X$ decays into 8--body final states or allow violation of
individual lepton numbers. Here we chose the second option, and
consider the following $X$ decay modes:

\begin{itemize}
\item 1--muon: $X \rightarrow \mu^- \bar{\nu}_{\mu} u \bar{d}$ \ \ or
  \ $X \rightarrow \mu^+ \nu_{\mu} \bar{u} d$ 
\item OS 2--muon: $X \rightarrow \mu^- \mu^+ \tau^- \tau^+$
\item SS 2--muon: $X \rightarrow \mu^- \mu^- \tau^+ \tau^+ $ \ or \ $X
  \rightarrow \mu^+ \mu^+ \tau^- \tau^-$
\item 4--muon:  $X \rightarrow \mu^- \mu^+ \mu^- \mu^+$
\end{itemize}

All these decays conserve total lepton number, but SS 2--muon decay
violates $\tau$ and $\mu$ number separately (as do $\nu_\mu
\leftrightarrow \nu_\tau$ oscillations). Note that we again use the
simple $X$ pair production cross section without Breit--Wigner factor
given in Eq.~(\ref{e1}) in this Subsection. Fitting the $X$ mass and
branching ratios to the in--cone multi--muon invariant mass and signed
multiplicity distributions, respectively, we find $m_X =
\unit[4.6]{GeV}$ and the branching ratios for the process $q\bar{q}
\rightarrow XX$ are \unit[82.32]{\%} for the 1--muon, \unit[5.99]{\%}
for the OS, \unit[10.25]{\%} for the SS 2--muon, and \unit[1.44]{\%}
for the 4--muon decay. Note that some of the secondary muons now come
from $\tau \rightarrow \mu$ decays, which produces rather soft
muons. Moreover, the phase space available for the two direct muons in
the $\mu\mu\tau\tau$ final states is quite small, i.e. these muons
tend to be rather soft as well, often failing the $p_T > 2$ GeV cut
applied on the secondary muons. As a result we need a significantly
larger 2--muon branching ratio than for the simple model.

\begin{figure}[htb]
\centering
\begin{minipage}[b]{8cm}
\includegraphics[width=8cm]{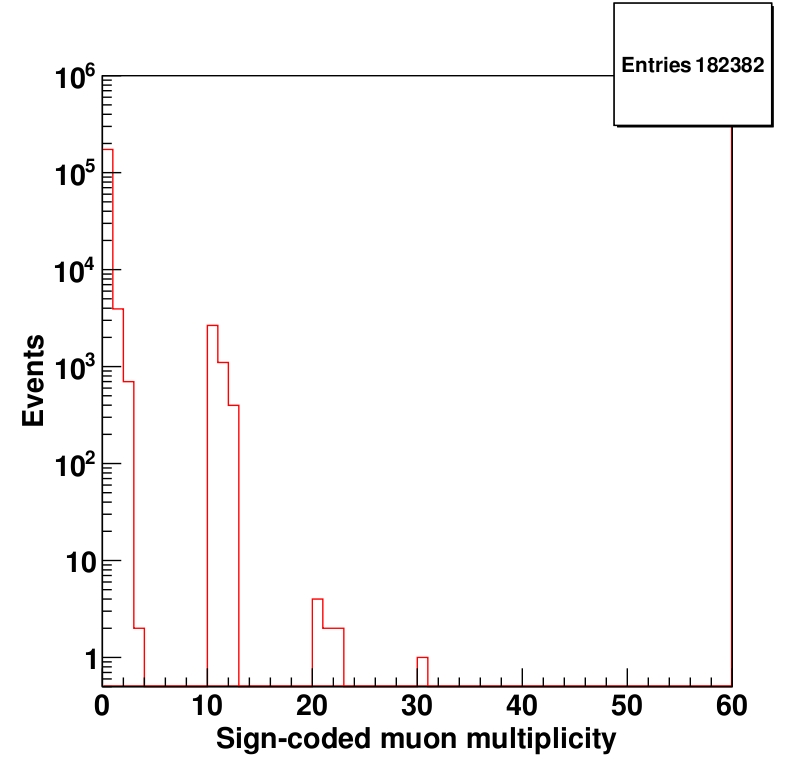}
\end{minipage}
\begin{minipage}[b]{8cm}
\includegraphics[width=8cm]{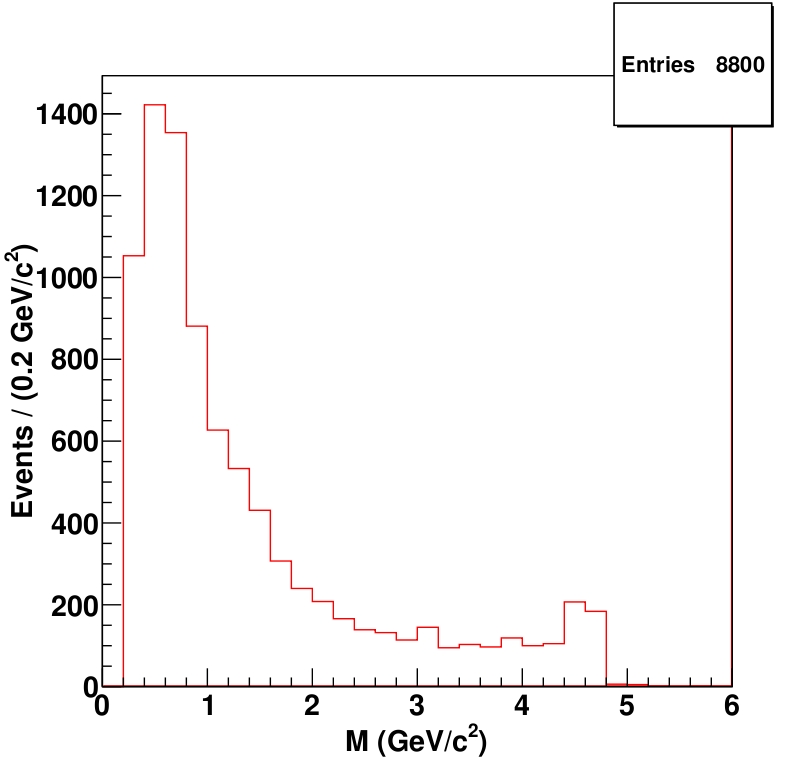}
    \end{minipage}
    \caption[]{Sign--coded muon multiplicity distribution of
      additional muons found in $36.8^\circ$ cones around initial
      muons (\textbf{left}) and invariant mass distribution of all
      muons within cones which contain at least one additional muon
      (\textbf{right}), as predicted by the model with $\mu$ and
      $\tau$ number violating $X$ decays. We assume $X$ pair
      production from $q\bar{q} \rightarrow XX$; 91,191 of 20 million
      simulated events pass the two--muon cuts.}
  \label{fig:Fig22bAnd34aTau}
\end{figure}

Fig.~\ref{fig:Fig22bAnd34aTau} shows the resulting distributions as
predicted using this modified model. Note that there is a small peak
at the tail end of the invariant mass distribution. This peak, which
is not observed in the CDF data, originates from the 4--muon decay of
$X$ particles, which is needed to reproduce the higher entries in the
multiplicity distribution.

The number of ghost events in various experiments predicted by this
model are listed in Table~\ref{tab:ResultsTau}. The prediction for UA1
is very similar to that of the simple model discussed in
Sec. \ref{sec:SM_Results}, see Table~\ref{tab:ResultsSM}. Mostly due
to the larger $X$ mass, the number of ghost events expected at the
fixed--target experiments is reduced, but at least for the $q \bar q$
production mechanism a significant number of events is again
predicted. 

\begin{table}
\centering
\caption[]{Expected number of ghost events for the model violating
  $\mu$ and $\tau$ number compared to the number of actually observed
  di--muon events. The normalization constants are $ N^{\tau}_{gg}
  \approx (1.18 \pm 0.06)\cdot 10^{-4}$ and $N^{\tau}_{q\bar{q}}
  \approx (2.1 \pm 0.1) \cdot 10^{-3}$. $X$ pair production from gluon
  fusion requires branching ratios $\unit[73.39]{\%}$ for the 1--muon,
  $\unit[2.99]{\%}$ for the OS, $\unit[22.56]{\%}$ for the SS 2--muon
  and $\unit[1.06]{\%}$ for the 4--muon decay.} 

\begin{tabular}{|l|l|l|l|l|}\hline
 & & \multicolumn{2}{c|}{\# ghost events for} & \# of observed \\
Exp. & $\int {\cal L} dt$ [Nucl./pb] & $gg \rightarrow XX$
& $q\bar{q} \rightarrow XX$ & $\mu\mu$ events \\\hline 
CDF & $114.41 \pm 6.51$ & $84895 \pm 4829$ & $84895 \pm 4829$ & $84895
\pm 4829$\\  
UA1--a & 0.108 & $1.4 \pm 0.1$ & $5.9 \pm 0.3$ & \multicolumn{1}{c|}{}\\
UA1--b & 0.6 & $10.5 \pm 0.6$ & $34.8 \pm 2.0$ & \raisebox{1.5ex}[-1.5ex]{880}\\
UA1--c & 4.7 & $82.0 \pm 4.7$ & $266.8 \pm 15.2$ & 2444\\
E605--a & $(1.14 \pm 0.08) \cdot 10^6$ & $2.0 \pm 0.2$ & $92.4 \pm
8.4$ & 43663 (OS)\\ 
E605--b & $(2.7 \pm 0.2) \cdot 10^5$ & $1.2 \pm 0.1$ & $60.1 \pm 5.7$
& 19470 (OS)\\ 
E772--a & $(5.8 \pm 0.3) \cdot 10^4$ & $6.6 \pm 0.5$ & $232.3 \pm
18.1$ & 83080 (OS)\\ 
E772--b & $3.5 \cdot 10^5$ & $59.9 \pm 3.5$ & $2002.8 \pm 116.4$ & $
\approx 450000 $ (OS)\\ 
E789 & $17.52 \pm 1.89$	& $2.7 \pm 0.3$	& $70.6 \pm 8.6$ & $ 71206 \pm
287 $ ($J/\psi$)\\ 
E866 & $3.78 \cdot 10^5$ & $14.3 \pm 0.8$ & $471.5 \pm 27.4$ & $
\approx 360000 $ (OS)\\ 
\hline
\end{tabular}
\label{tab:ResultsTau}
\end{table}

\subsection{Combined Model}

The combination of the Breit--Wigner resonance in the $X$ pair
production cross section with the $\tau$ and $\mu$ number violating
$X$ decay modes enables us to reproduce the main characteristics of
the observed ghost events. As in the previous Subsection we need $m_X
= \unit[4.6]{GeV}$ to reproduce the in--cone multi--muon invariant
mass distribution.

We saw above that the muons from $\tau$ decays and the directly
produced muons in the two--muon decay modes are rather soft in this
decay scenario. In order to reproduce the observed gradual decline of
the multi--muon invariant mass distribution we therefore require
larger values of $m_Y$ than in the scenario where $X$ decays conserve
all lepton numbers separately. Specifically, if all $X$ pairs are
produced from $q \bar q$ annihilation, we need $m_Y = \Gamma_Y =
\unit[110]{GeV}$, whereas gluon fusion dominance requires $m_Y =
\Gamma_Y = \unit[180]{GeV}$.

This in turn increases the number of $X$ particles produced with large
$p_T$, and hence the efficiency with which additional muons pass the
$p_T > 2$ GeV cut. The branching ratios for the multi--muon channels
therefore have to be adjusted downward relative to the model without
Breit--Wigner factor. In case of $q \bar q$ production, we find
\unit[91.01]{\%} for the 1--muon, \unit[3.76]{\%} for the OS,
\unit[2.86]{\%} for the SS 2--muon, and \unit[2.37]{\%} for the
4--muon decay. In case of gluon fusion the branching ratios are
\unit[91.91]{\%} for the 1--muon, \unit[4.22]{\%} for the OS,
\unit[3.25]{\%} for the SS 2--muon, and \unit[0.62]{\%} for the
4--muon decay.

\begin{figure}[htb]
  \centering
	\begin{minipage}[b]{8cm}
	\includegraphics[width=8cm]{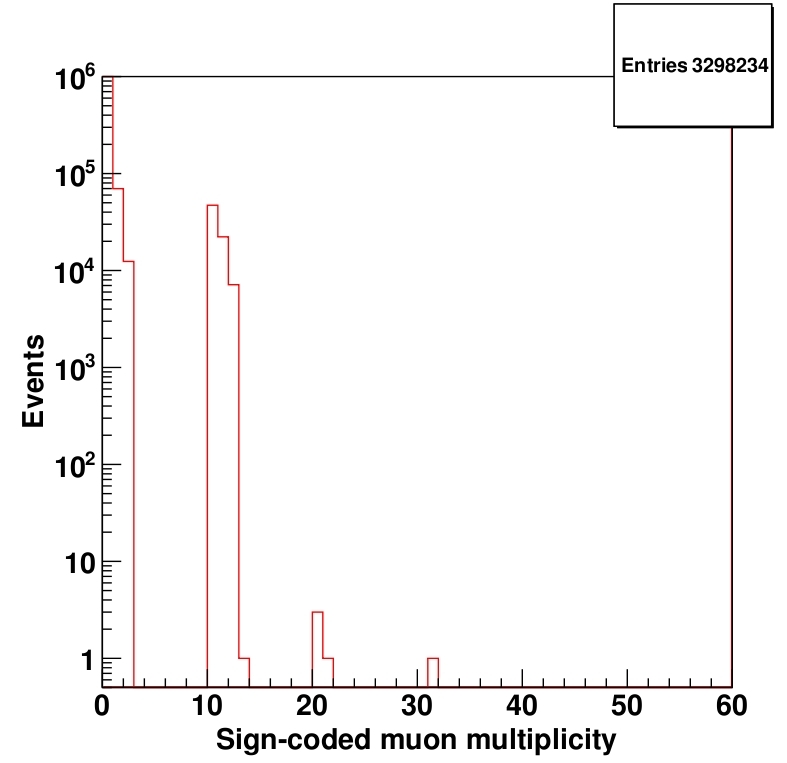}
        \end{minipage}
    \begin{minipage}[b]{8cm}
    \includegraphics[width=8cm]{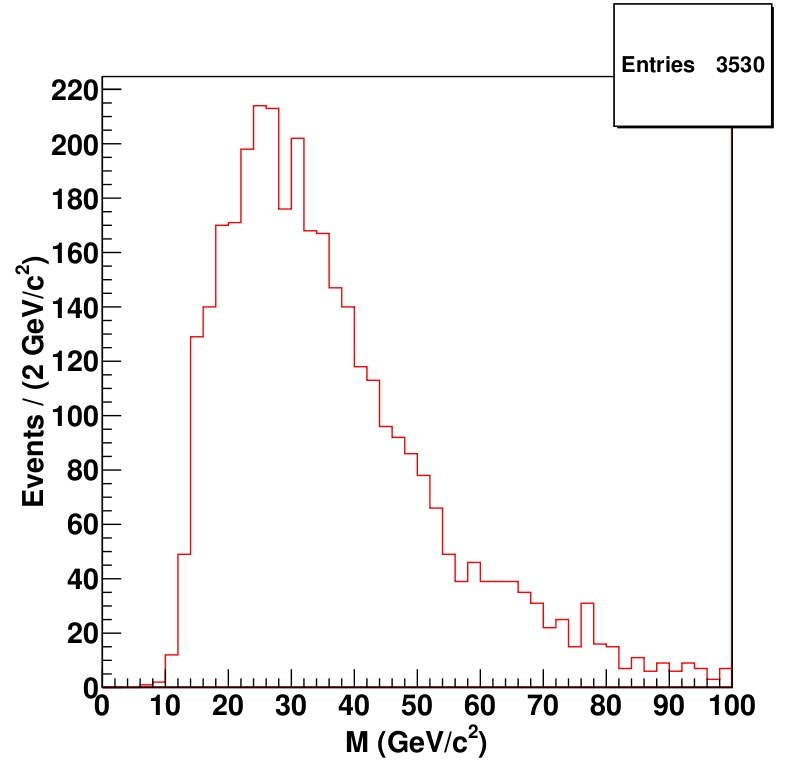}
    \end{minipage}
  \caption[]{Sign--coded muon multiplicity distribution of additional
    muons found in $36.8^\circ$ cones around initial muons
    (\textbf{left}) and invariant mass distribution of all muons
    within events in which both cones contain each at least one
    additional muon (\textbf{right}), for $gg \rightarrow Y
    \rightarrow XX$ with the $\tau$ and $\mu$ number violating $X$
    decay modes. 1,649,117 of 20 million generated events pass the
    two--muon cuts.} 
  \label{fig:Fig22bAnd35aTauBW}
\end{figure}

\begin{figure}[htb]
  \centering
  \includegraphics[width=8.2cm]{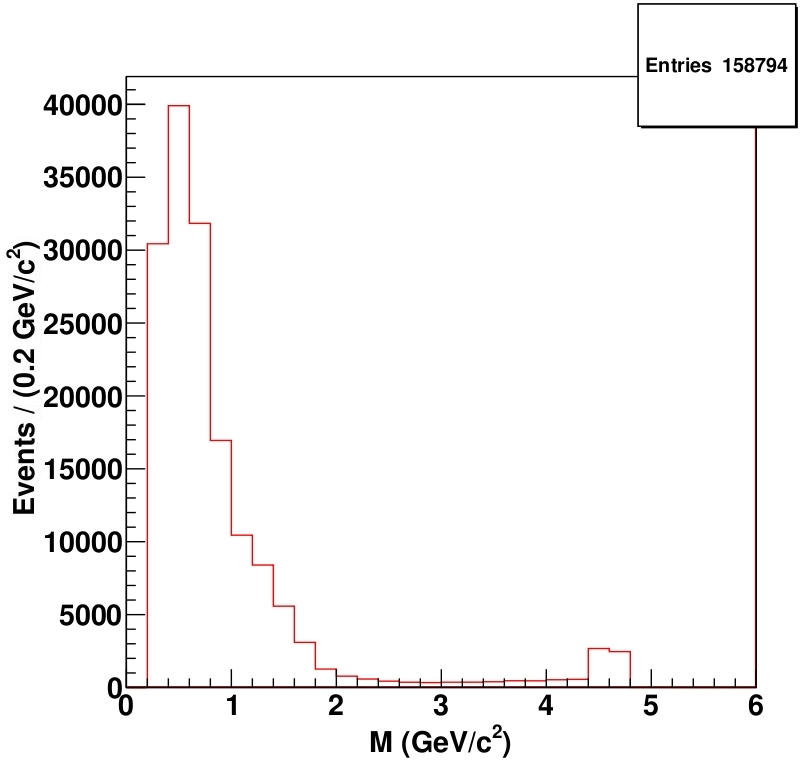}
\caption[]{Simulated invariant mass distribution of all muons within
  cones which contain at least one additional muon, using $gg
  \rightarrow Y \rightarrow XX$ with $\mu$ and $\tau$ number violating
  $X$ decays. 1,649,117 of 20 million generated events pass the
  two--muon cuts.} 
  \label{fig:Fig34aTauBW}
\end{figure}

There is a better reproduction of the muon multiplicity distribution
for the gluon fusion compared to the quark--antiquark
annihilation. Not only the entries in ``1'' and ``10'' are reproduced
well, also the cones with two and three additional muons are in closer
agreement with the measurement. The simulation results are shown in
Fig.~\ref{fig:Fig22bAnd35aTauBW}. In addition, the nearly complete
absence of the 4--muon decay has the desirable side--effect that the
peak at the mass of the $X$ particle in the simulation of Fig.~34a
almost vanishes; see Fig.~\ref{fig:Fig34aTauBW}.

In this model around 25 (200) ghosts should have appeared at UA1 for
gluon fusion ($q \bar q$ annihilation).  As for the Breit-Wigner
resonance with lepton number conserving $X$ decays, we do not expect
ghost events in the data samples of the fixed--target
experiments. 

Recall that the poor acceptance of these experiments is a result of
the cuts tailored for analyses of Drell--Yan production. If this cut
can be relaxed, a significant number of ghost events could be detected
even at fixed--target energies. In particular, a hypothetical
fixed--target experiment with a center of mass energy of
$\unit[38.8]{GeV}$ and an integrated luminosity of $\unit[1 \cdot
10^5]{Nucl./pb}$ could probe the Breit--Wigner model with $\mu$ and
$\tau$ number violating $X$ decay modes. If the initial muons each
have a lab energy $E_{\mu} \geq \unit[5]{GeV}$, we would expect
$3276.0 \pm 190.4$ ($408.9 \pm 23.8$) ghost events for
quark--antiquark annihilation (gluon fusion), if full angular
acceptance is assumed. Hence, a fixed--target experiment with
good spatial coverage can test even this most difficult model
decisively, if the vertex resolution is sufficient to detect the decay
length of the $X$ particles.

\section{Summary and Conclusions}

In this paper we investigated the question whether CDF ``ghost''
events could have been detected by experiments operating at lower
energies. The answer to this question depends on the model chosen to
describe the CDF data. We started by constructing a simple model, in
which two light $X$ particles are produced in an S--wave, and decay
into four--body final states containing at least one muon, with a
decay length of 20 mm. As far as we can tell, this model describes the
di--muon ghost sample fairly well; recall that this accounts for more
than 90\% of all ghost events. Since the $X$ particles are light, they
could be produced not only at the CERN SpS collider, but also in
fixed--target collisions at Fermilab. The latter should have
accumulated hundreds to thousands of such events in their data,
depending on whether $X$ pair production is dominated by gluon fusion
or by $q \bar q$ annihilation. In half of these events both muons
should have the same charge; there is very little physics background
for such like--sign pairs at fixed--target energies, where $b \bar b$
production is suppressed (which can lead to like--sign pairs via $B^0
- \overline{B^0}$ oscillations). Unfortunately the published analyses
of di--muon data from these experiments are based on triggers that
require the presence of two opposite--sign muons in the event. Some
experiments evidently also recorded same--sign pairs, but the
effective luminosities and efficiencies for these are not stated. On
the other hand, at least one of the experiments, Fermilab E789, should
have had no trouble detecting the typical $X$ decay length of several
cm.

However, this simple model does not reproduce the ghost events with
more than two muons very well. It does not predict sufficiently many
secondary muons with the same charge as the primary one, and it
predicts a too steep fall--off of the multi--muon invariant mass
distribution. The first problem can be cured by allowing $X$ decays to
violate $\mu$ and $\tau$ number, still respecting the overall lepton
number, while the second problem can be solved by assuming that $X$
pair production proceeds via a very broad resonance $Y$. This more
complicated model still predicts that the UA1 di--muon sample should
contain an appreciable number of ghost events; however, these would be
overwhelmed by di--muon events from conventional sources, in
particular from $b \bar b$ production, unless the long decay length of
the $X$ particles could be detected. Moreover, this model predicts
that few or no ghost events should be contained in the Fermilab fixed--target di--muon data. However, these data were subjected to angular
cuts that isolate Drell--Yan events, but are very inefficient for
ghost events. We saw that a fixed--target experiment with good angular
coverage and sufficient tracking resolution to detect the finite $X$
decay length should be able to decisively probe even this more complex
model. 

Given the spectacular nature of the ghost events, it seems unlikely to
us that they would have escaped detection, had they been produced at
rates similar to those predicted by our simple model. However, a
proper analysis can only be performed by the experimental groups
themselves; this is true in particular in view of the apparently quite
complicated acceptances of these experiments. We hope that this paper,
and the ghost models we constructed and embedded into HERWIG++, will
facilitate such analyses.

\subsection*{Acknowledgments}
NB wants to thank the ``Bonn-Cologne Graduate School of Physics and
Astronomy'' and the ``Universit\"at Bonn'' for financial support. 

\begin{appendix}

\end{appendix}


\begin{thebibliography}{500}
 
\bibitem{Ghost}
T. Aaltonen \textit{et. al.},
arXiv:0810.5357v2 [hep-ex].

\bibitem{Herwig}
M. B\"ahr \textit{et. al.},
Eur. Phys. J. C \textbf{58}, 639 (2008), arXiv:0803.0883 [hep-ph].

\bibitem{UA1}
C. Albajar \textit{et al.},
Phys. Lett. B \textbf{186}, 237 (1987);
C. Albajar \textit{et al.},
Phys. Lett. B \textbf{256}, 121 (1991).

\bibitem{E605}
G. Moreno \textit{et. al.},
Phys. Rev. D \textbf{43}, 2815 (1991).

\bibitem{E772}
P. L. McGaughey \textit{et. al.},
Phys. Rev. D \textbf{50}, 3038 (1994);
P. L. McGaughey \textit{et. al.},
Phys. Rev. D \textbf{60}, 119903 (1999);
D. M. Alde \textit{et. al.},
Phys. Rev. Lett. \textbf{64}, 2479 (1990).

\bibitem{E789}
D. M. Janson \textit{et. al.},
Phys. Rev. Lett. \textbf{74}, 3118 (1995).

\bibitem{E866}
E. A. Hawker \textit{et. al.},
Phys. Rev. Lett. \textbf{80}, 3715 (1998);
R. S. Towell \textit{et. al.},
Phys. Rev. D \textbf{64}, 052002 (2001).

\bibitem{mueller}
T. M\"uller, Nucl. Instr. Meth. A \textbf{252}, 387 (1986).

\end{thebibliography}
\end{document}